\documentclass[twocolumn,showpacs,aps,amsmath,amssymb,pre,superscriptaddress,nofootinbib]{revtex4-1}

\usepackage{graphicx}
\usepackage{amssymb}
\usepackage{amsmath}
\usepackage{epstopdf}
\usepackage{color}

\begin{document}
\title{The complex dynamics of memristive circuits:\\ analytical results and universal slow relaxation}

\author{F. Caravelli}
\affiliation{Invenia Labs, 27 Parkside Place, Parkside, Cambridge CB1 1HQ, UK}
\affiliation{London Institute for Mathematical Sciences, 35a South Street, London W1K 2XF, UK}

\author{F. L. Traversa}
\affiliation{Department of Physics, University of California, San Diego, La Jolla, CA 92093, USA}
\author{M. Di Ventra}
\affiliation{Department of Physics, University of California, San Diego, La Jolla, CA 92093, USA}

\begin{abstract}
Networks with memristive elements (resistors with memory) are being explored for a variety of applications ranging from unconventional computing to  models of the brain. However, analytical results that highlight the role of the graph connectivity on the memory dynamics are still a few, thus limiting our understanding of these important dynamical 
systems.  In this paper, we derive an exact matrix equation of motion that takes into account all the network constraints of a purely memristive circuit, and we employ it to derive analytical results regarding its relaxation properties. We are able to describe the memory evolution in terms of orthogonal projection operators onto the subspace of fundamental 
loop space of the underlying circuit. This orthogonal projection explicitly reveals the coupling between the spatial and temporal sectors of the memristive circuits and compactly describes the circuit topology. For the case of disordered graphs, we are able to explain the emergence of a power law relaxation as a superposition of exponential relaxation times with a broad range of scales using random matrices. This power law is also {\it universal}, namely independent of the topology of the underlying graph but dependent only on the density of loops. In the case of circuits 
subject to alternating voltage instead, we are able to obtain an approximate solution of the dynamics, which is tested against a specific network topology. These result suggest a much richer dynamics of memristive networks than previously considered. 
\end{abstract}


\maketitle
\section{Introduction}
The role of memory in the statistical properties of complex systems is emerging as an important new direction of study \cite{Verstergaard,Caravelli2015,Caravelli2016}. In particular, 
memristive circuits (circuits made of resistors with memory) are attracting considerable attention in view of their similarities with the dynamics of self-organizing systems such as swarms (e.g., ants), and their ability to solve certain optimization problems \cite{Dorigo,traversa14b,Traversa2014,Traversa2015}. In fact, the physical properties of memristors are relevant both for their practical use, 
such as in the field of unconventional computing \cite{indiveri,traversa13a, Avizienis,Stieg12}, as well as to understand the collective behavior and learning abilities of certain biological systems \cite{pershin09b,diventra13a,pershin11d,festchua,pershin13b}, including the brain \cite{chialvo,pershin10c}. A key signature of these networks is the presence of time nonlocality (memory), a feature, that coupled with Kirchhoff's conservation laws, promotes unexpected phenomena, such as first-order phase transitions as a function of memory content \cite{Driscoli,PershinPT} or avalanches \cite{Sheldon}.

Yet, very little analytical advances have been made for complex circuits made of memristors, due to the strong non-local behaviour introduced by the network constraints, such as the circuit conservation laws, that make numerical results all but necessary. In turn, it is still unclear how the memory dynamics of each element depends on the graph connectivity. Another important issue is the role of memory in the {\it relaxation to steady state} of memristive networks. 
In other words, the question ``How does an excitation in a disordered network of memristive elements relax to steady state ?'', has yet to be answered. As mentioned above, this is not just an academic exercise: these types of networks are being employed to solve complex problems in a variety of different modes. Hence, an answer to this query bears immediate relevance to the question of how efficient such systems are as computing machines and how fast they converge to the their asymptotic stable states. 

In this paper we make three fundamental advances. For the case of linear current-controlled memristors, we demonstrate a closed matrix equation of motion for the \textit{internal} memory of the circuit, which embeds the conservation laws of the system. With this equation in hand, we show that 
such networks can support scale-free temporal correlations induced by the network non-local properties. We provide an analytical demonstration of this fact using the simplest model of linear memristors, which is a  good approximation for a variety of actual physical systems \cite{stru12, strukov08, strukov05a, chua76a, diventra09a, demin15}. In fact, by means of graph-theoretic tools we show explicitly that the spatial and temporal sectors of the dynamics are coupled by orthogonal projections onto the subspace of fundamental 
current loops. This coupling ensures the emergence of a power law as superposition of exponential relaxations times with a broad range of scales, which is the typical signature of ``glassy" behavior. 
Slow relaxation phenomena have been already observed experimentally. Specifically, in \cite{Avizienis,Stieg12} it has been observed that the frequency spectrum of the resistance in atomic switch networks is a power law, which could be due either to self-organized criticality or to a superposition of a broad range of relaxation time-scales as we observe in the present paper. 

Ultimately, our derived equation may serve as the basis for further analysis of the relaxation properties of circuits with memory. In fact, for the case of AC forcing, we are able to obtain for the first time an approximate analytical solution in the case in which the projector operator is diagonally dominant. We test our obtained approximate solution in the case of a specific graph configuration. 

\section{Methods}
Let us start by considering the exact solution of a linear circuit~\footnote{In this paper we consider only linear relations between voltages and currents.}, written in terms of graph quantities such as the loop matrix description of a circuit \cite{nilsson}, focusing on linear memristors. 
Specifically, we employ a slightly modified version of a widely used linear model of memristors described by the equations that relate the current $I(t)$ to the voltage $V(t)$ \cite{strukov08}:
\begin{eqnarray}
V(t) &=& R(w,t) I(t) \\
\dot w &=& \frac{{\mathcal J}}{\beta} R_{on} I + \alpha w\,,
\end{eqnarray}
where $w$ is the internal memory state variable, ${\mathcal J}=\pm1$ represents the polarity of the memristor, $R_{on}$ the limiting resistance when the memristor is in the conducting phase, and $\beta$ is a constant. In the case of the memristors of Ref.~\cite{strukov08}, made of an oxide thin film sandwiched between two metal layers with oxygen vacancies, one has $\beta=\frac{2 d^2}{\mu}$, where $\mu$ represents the electron mobility and $d$ the size of the memristor. The parameter $\alpha$ quantifies the rate of decay of the memory when all generators are 
switched off. The memory resistance we consider is limited between the values $R_{off}$ in the insulating phase and $R_{on}$ in the conducting one, and
depends linearly only on the dynamical internal parameter $w(t)$, 
\begin{eqnarray}
R(w,t)&=&R_{on} \left(1-w(t)\right)+R_{off} w(t) \nonumber \\
&=& R_{on}[1+(r-1) w(t)],
\label{eq:memR}
\end{eqnarray}
where we have implicitly defined the constant $r=R_{off}/R_{on}$, typically $r\gg 1$. 

\begin{figure}
\centering
\includegraphics[scale=0.33]{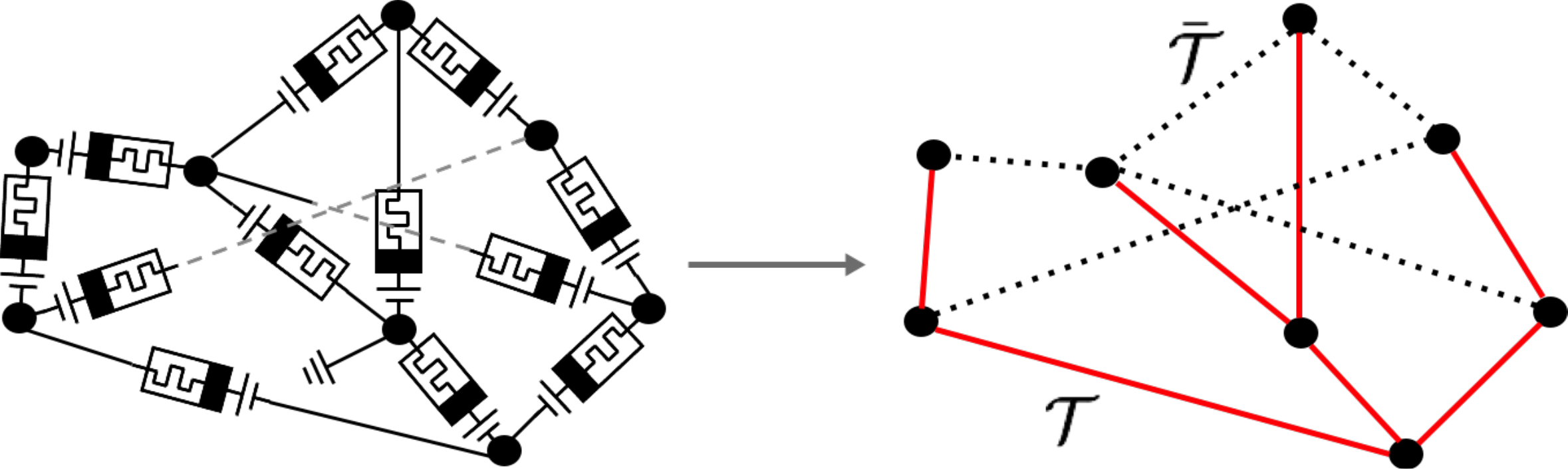}
\caption{Small instance of a random memristive network considered in this work with its chord and co-chord decomposition depicted on the right. The formal solutions of the currents can be written in terms of the fundamental loops of the circuit. Given a circuit and an orientation of the currents, we first find a spanning tree $\mathcal T$, which in the figure is given by the red edges. Each element of the tree, $\mathcal T$ is a \textit{chord}. Every remaining edge which is not in the spanning tree set is called \textit{co-chord}; for each \textit{co-chord} it is possible to assign a \textit{mesh} variable or \textit{fundamental loop.} 
}
\label{fig:network}
\end{figure}

For generic linear circuits, it is well known that one can write the solution of the current configuration as a function of the current and voltage sources and the {\it cycle matrix} $A$ of the graph associated to the circuit \cite{nilsson}. In order to understand the derivation in simple terms,  consider Fig. \ref{fig:network}. The cycle matrix is a rectangular matrix of size $L\times M$, where $L$ is the number of {\it fundamental loops} and $M$ the number of resistors/memristors. Its introduction is motivated by the following observation: due to the Kirchhoff's constraints on the currents, only a certain number of currents -- which equals the number of fundamental loops of the circuit -- are linearly independent. The number of fundamental loops can be easily calculated from basic graph theory \cite{bollobas}, $L=M-N+1$, where $N$ is the number of nodes of the circuit, and $N-1$ is the number of edges in the tree $\mathcal T$, called chords. The complementary set of edges is denoted with $\bar {\mathcal T}$, and these edges called co-chords. Therefore, the number of fundamental loops is equal to the number of co-chords. 

To be specific, let us consider the case in which there are no current sources, only voltage sources parametrized as elements of a vector $\vec S(t)$ on the set of edges (or arcs) of the graph. Similarly, let us introduce a diagonal matrix of (mem)ristances $R=\text{diag}(R_i)$, where the index $i$ runs over the edges of the network. The formal solution of the current configuration, $\vec i$, as a function of $R$, $\vec S$ and $A$ is then given by \cite{nilsson}:
\begin{equation}
\vec i=- A^t \left( A R A^t\right)^{-1} A \vec S(t).
\label{eq:init}
\end{equation}

The derivation is standard but elegant and is provided for completeness in the Supplemental Material.
Equation \eqref{eq:init} is the starting point of our analysis. We 
consider the physically relevant case of a decay to the $R_{off}$ state when there are no sources in the circuit. This is consistent with experimental observations (see, e.g., \cite{Stieg12}). Given a diagonal matrix $P$ such that $P=\text{diag}(\sigma_1,\cdots,\sigma_M)$, where $\sigma_j$ is 1 for all memristors up to $M$, one has $\bar R=P R$, where $R$ is still a positive diagonal matrix and contains the absolute values of the resistances. We can however absorb the matrix $P$ into the matrix $A$,  by defining $\bar A= A P$.

We now note that the resistance matrix is the one of linear memristors as in \eqref{eq:memR}, where we introduce the internal memory vector $\vec W=\{w_i\}$, i.e., 
\begin{equation}
	R=\text{diag}(R_{on}(\vec 1+(r-1)\vec W(t))).
\end{equation}
If we introduce the diagonal matrix $W=\text{diag}(\vec W)$,
we can then write the equation for the internal memory states as the following equation 
\begin{eqnarray}
 \frac{d \vec W}{d t}&=&\alpha \vec W \nonumber \\
&-&\frac{1}{\beta} \mathcal J A^t \left( \bar A A^t+(r-1)\bar A  W A^t \right)^{-1} A \vec S( t),\nonumber \\ 
\label{eq:dynnn2}
\end{eqnarray}
where  $\mathcal J$ is the matrix which contains the polarity of the memristors. For $\alpha>0$ the resistance decays to the $R_{off}$ state in the absence of sources, and for $\alpha<0$ to the $R_{on}$ state. This term is independent from the cycle matrix $A$, meaning that this is a property of each single memristor, and not a global network effect. We consider the case of homogeneous memristor properties, i.e., they all have identical off and on states. It is of course easy to generalize our results to the inhomogeneous case.

These equations are insofar general. For any circuit, the inverse of $\bar A A^t+(r-1) \bar A  W A^t$ exists so long as $R$ has all non-zero entries on the diagonal, which is the case if both $R_{on}$ and $R_{off}$ are either positive or negative. 
In order to simplify the notation, we introduce the matrix $\bar \Omega = A^t (\bar A A^t)^{-1} \bar A$, and $\bar S=P \bar S$. By construction, $\bar \Omega$ is an orthogonal projector onto the subspace of fundamental current loops if all resistances in the circuit are all either positive or negative, while it is non-orthogonal 
there is a mixture of positive and negative resistances.

Let us now set $\xi=r-1$. After a lengthy but trivial computation we then derive the following equation for the internal memory (see Supplemental 
Material):
\begin{eqnarray}
 \frac{d \vec W}{d t} &=&\alpha \vec W- \frac{1}{\beta}\mathcal J (\hat I+\xi\ \bar\Omega  W)^{-1} \bar \Omega \bar { S}( t), 
\label{eq:dynnn5}
\end{eqnarray}
with $\hat I$ the identity matrix. 
 This is the central result of our paper. It is a compact equation that describes the dynamics of the internal memory states of memristors in linear circuits based upon projection operators.

Few comments are in order. First of all,  \eqref{eq:dynnn5} has been derived with the assumption of invertibility of $W$. Strictly speaking, this 
means that we are considering the bulk of the dynamics, i.e., when no memristor is in the $R_{on}$ state. Nevertheless, the final formula is independent of $W^{-1}$ and is numerically well-behaved for $w_i\approx 0$, which suggests it can be extended to the boundaries as well. We note moreover that $\xi$ plays the role of the amount of non-linearity in the systems.

In addition, \eqref{eq:dynnn5} satisfies all the network constraints and Kirchhoff's laws. The importance of the number of fundamental loops is shown by the fact that $\text{dim(Span($\bar \Omega$))}=M-N+1\equiv L$, which implies that the operator $\bar \Omega$ contains information only on the fundamental loops of the circuit. %

We have that $P\rightarrow \hat I$, and thus $\bar  \Omega\rightarrow\Omega$, where $\Omega=A^t (A A^t)^{-1} A$ is an orthogonal projection. This implies that we can always decompose any matrix or vector $R=\Omega R+(\hat I-\Omega) R=R_\Omega+\tilde R$, with $\Omega R=R_\Omega$. For the case of passive components,
 we can identify the operator $\Gamma=(I-\Omega)$ as $B(B^t B)^{-1} B^t$, with $B$ being the reduced incidence matrix (see Supplemental Material).
 
Given the matrix $(A\ B^t)$, we can write the identity $I=(A\ B^t) (A\ B^t)^{-1}$ and using the fact that  $B^t A=0$, it is easy to prove that $I=A^t(A A^t)^{-1} A+B(B^t B)^{-1} B^t=\Omega+\Gamma$, which provides a nice interpretation for the complementary projector. Moreover, the separation between linearity and non-linearity is explicit in  \eqref{eq:dynnn5} and is controlled by the constant $r-1=(R_{off}-R_{on})/R_{on}$.

We now study the consequences of \eqref{eq:dynnn5} numerically focusing on passive elements only. We take advantage of the fact that there is a simple parametrization for projector operators, given that we are interested in the \textit{average} properties of the dynamics. In this way, the only two relevant parameters are $M$ and $N$, the number of memristors and the number of nodes, respectively.  We then generate a 
random matrix $A$ of size $L \times M$ 
and evaluate the matrix $\bar \Omega$ according to the equation $ \Omega=A^t ( A A^t)^{-1}  A$. The matrix $A$ is of the form $A=( I\ A_{\tau})$, where $A_{\tau}$ is generated using random entries with probability $1/3$ for the discrete values $\{-1,0,1\}$. 
We then consider the quenched dynamics for the memory parameters $w_i$ by integrating  \eqref{eq:dynnn5} numerically using an explicit Euler method and studying the relaxation to steady state. In addition to the bulk equation, we also included the constraints on the internal memory states $0<w_i<1$. 

\section{Results}
For the present paper we focus on the case without active components, i.e., we set $P=\hat I$. For the case of constant-voltage relaxation, we study the relaxation numerically, and using  \eqref{eq:dynnn5} we provide arguments to explain its average behavior.

\subsection{DC relaxation}
 
\begin{figure}[ht!]
\centering
\includegraphics[scale=0.3]{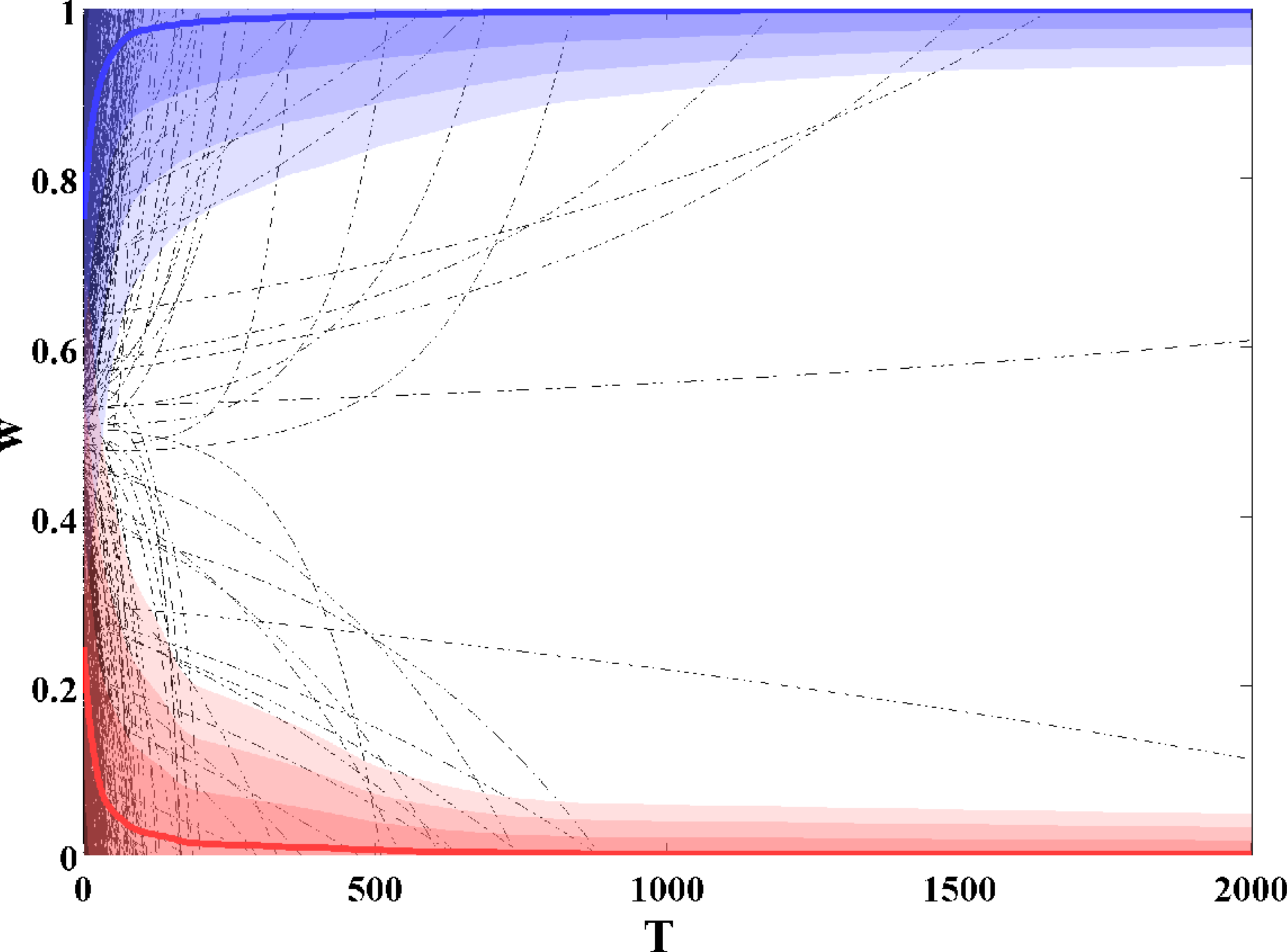}\\
\includegraphics[scale=0.3]{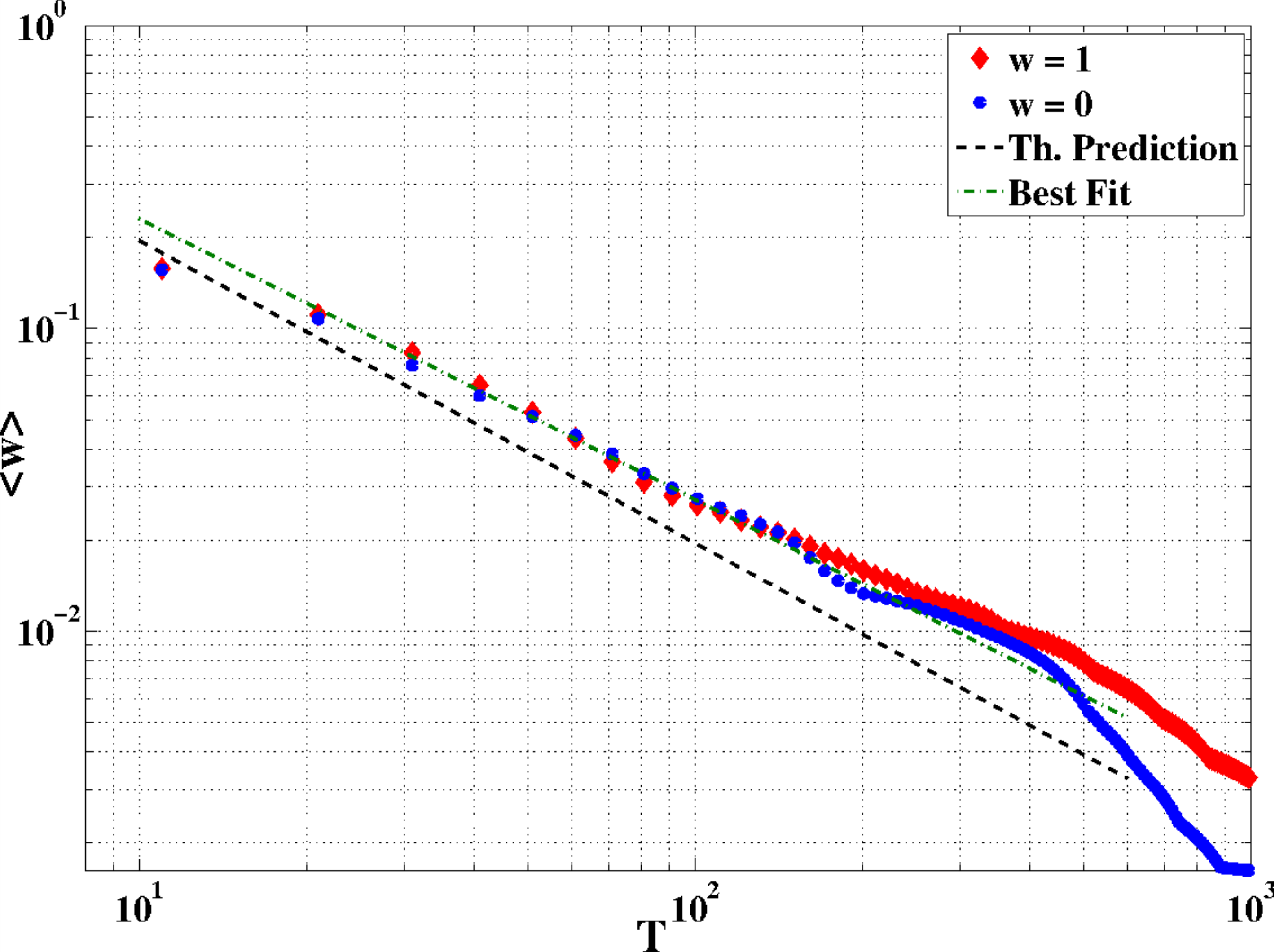}
\caption{Average thermalization of the internal memristor memory as a function of time, made assuming that  $M=600$, $L=300$ for input voltages chosen at random in $[-5,5]$ Volts and $\beta=10^{-1}$. \textit{Top}: Simulation of  \eqref{eq:dynnn5} for a single instance of the randomly selected projector for each memristor (black dashed curves) and their means for those approaching $w=1$ limit (blue curve) and $w_i=0$  (red curve). The shading represents the error at 1, 2 and 3 $\sigma$ to show the sensitivity of the relaxation. \textit{Bottom:} Best fits in the power law regime of the average memory parameters as a function of time for the $w_i=0$ (blue) limit and $w_i=1$ (in red, where we plot $1-\langle w\rangle$) and the best fit (black dashed curve). We observe a relaxation behavior which is compatible with a power law $\langle w(t)\rangle\approx t^{-\rho}$, with a best fit exponent of $\rho\approx 0.93$, against the $\rho=1$ predicted theoretically.}
\label{fig:secondmod}
\end{figure}

We consider first the case in which the applied voltage is constant in time. Such case differs from the AC one, since every memristor eventually reaches an asymptotic boundary value of 0 or 1. We performed numerical simulations for the evolution of the average memory parameters $\langle w \rangle$ as a function of time for each single realization in Fig. \ref{fig:secondmod} (top).  We observe that the relaxation behavior is characterized by a slow convergence towards the asymptotic values. We differentiate the trajectories reaching the 1's values from the ones reaching the 0's values. In Fig. \ref{fig:secondmod} (top) we show that both trajectories can be fitted by a power law. We observe this in the limit $r\gg 1$ and in the numerics we choose $r\approx1000$. The blue and red curves represent the average parameters for the superior and inferior boundary of the memory,  respectively. 
In Fig. \ref{fig:secondmod} (bottom) we plot the best fit with power law (red curve) $\langle w(t) \rangle \approx t^{-\rho}$, with $\rho\approx 0.93$, where the top boundary has been inverted, i.e.,  $\langle w\rangle^\prime=1-\langle w \rangle$. (Finite size effects are discussed in the Supporting Information.)

Let us approximate the curve of each memristor as an exponential, $w_i(t)=w^0_i e^{-\lambda_i t}$, where $w^0_i$ is the initial value of the $i$th memristor. We focus only on the memristors which converge to $w_i=0$ values. 
We then assume that there exists a certain distribution $P(\lambda)$ of decay times, $\lambda_i$. 

The average behavior of the internal memory is thus given by: 
\begin{equation}
\langle w_i(t) \rangle\approx \langle w_{i}^0 \rangle \int e^{-\lambda t} P(\lambda) d\lambda
\end{equation}
where we assumed that $w_{i}^0$ and $\lambda_i$ are uncorrelated.  If we introduce the inverse time scale $\tilde \lambda$, $P(\lambda)=\tilde \lambda e^{-\frac{\lambda}{\tilde \lambda}}$, one has 
\begin{equation}
\langle w_i(t) \rangle=\frac{\tilde \lambda}{2} \int_0^\infty e^{-\frac{\lambda}{\tilde \lambda}} e^{-\lambda t} d\lambda=  \frac{1}{2}\frac{1}{1+\tilde \lambda t}
\label{eq:approx}
\end{equation}
where $\tilde \lambda$ is an artificial time scale, playing the role of a cutoff. We thus obtain the result that for $t\gg \frac{1}{\tilde \lambda}$, on average, memristors thermalize to the steady state as $\approx 1/t$. 

\begin{figure}
\centering
\includegraphics[scale=0.3]{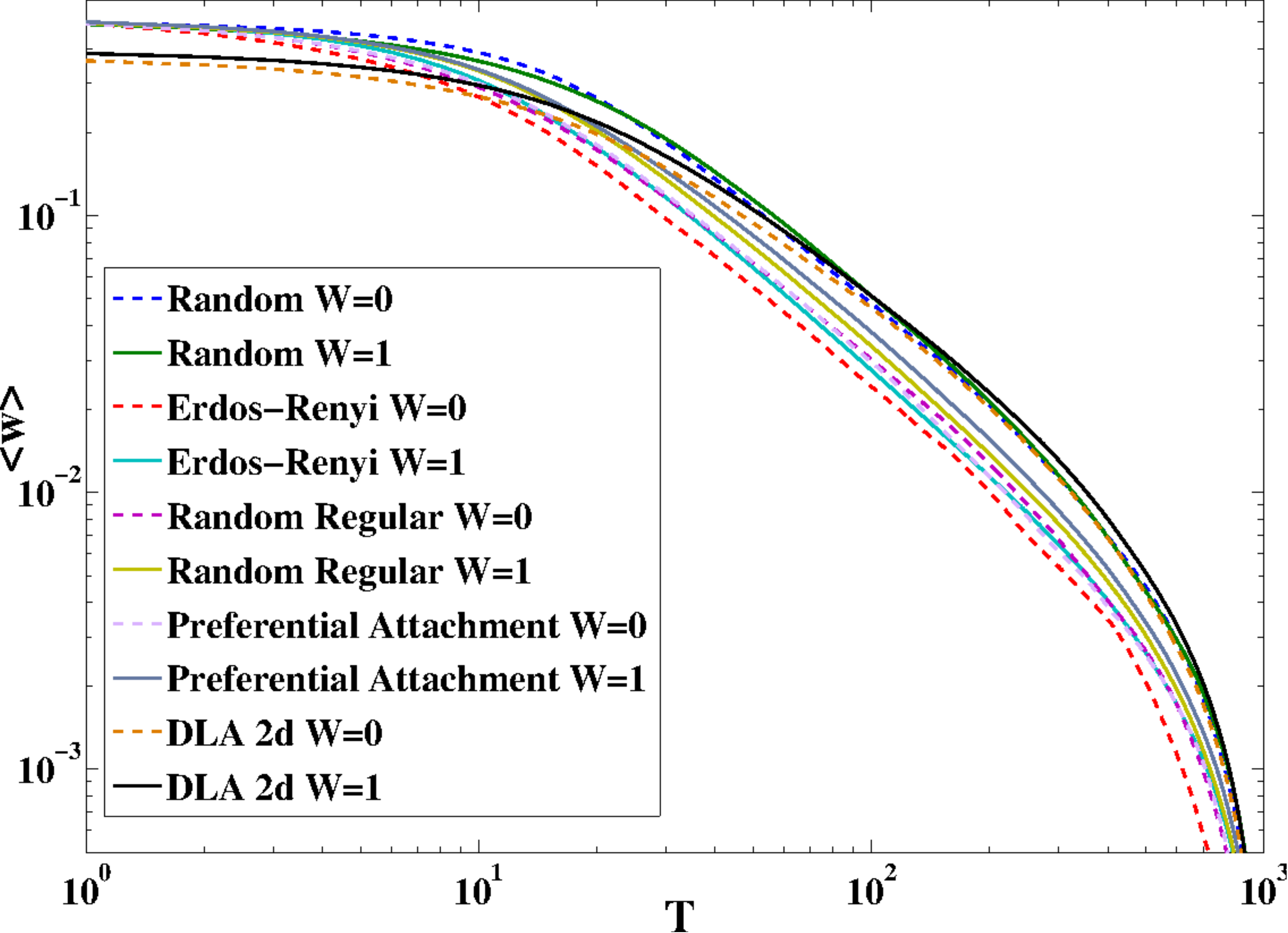}
\caption{The relaxation of the average internal memory for various memristive circuit topologies. We consider the case of random loop matrices, Erdos-Renyi graphs, Random regular graphs, Preferential Attachment graphs and Diffusion-Limited Aggregation in 2 dimensions. Results are averaged over 100 random graphs for each class, with approximately 400 memristors.}
\label{fig:allgraphs}
\end{figure}

 Using the relation between the projector in the loop space and the complementary projector based on the incidence matrix, it is easy to simulate arbitrary topologies. In fact, we note that such 
 ``glassy'' behaviour shows some universal properties, i.e., {\it it is independent of the graph topology} used. We show this explicatively in Fig. \ref{fig:allgraphs} where we plot 
the network relaxation for various topologies such as the Erdos-Renyi graphs, Random Regular Graphs, Preferential Attachment graphs and Diffusion-Limited Aggregation graphs. The relaxation behaviour 
observed is practically identical in all case. Such behavior then suggests a {\it universal} property of these networks.

\begin{figure}
\centering
\includegraphics[scale=0.32]{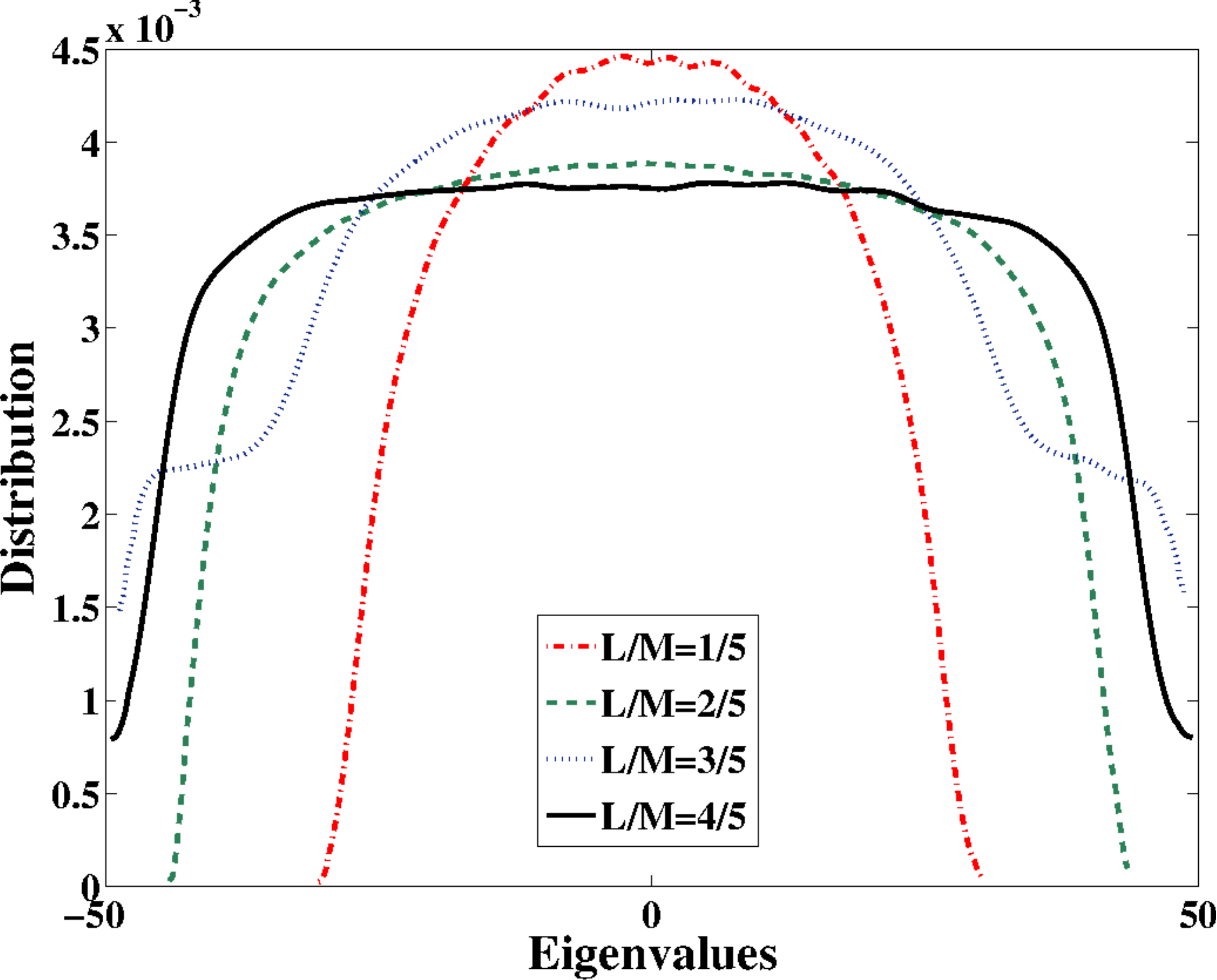}
\caption{Distribution of eigenvalues of the matrix $O$ (from \eqref{eq:dynnndc}) for the case $\alpha=0$, using the class of random loop matrices $A$ and for random voltage vectors in the interval $S=[-50,50]$, and averaged over 500 iterations, where the zero eigenvalues have been removed. We plot $L/M=1/5,2/5,3/5$ and $4/5$ with $M=500$. The distributions have been smoothed using gaussian kernels, and are zero outside the specified range.}
 \label{fig:eigs}
 \end{figure}

In order to understand the relation between the spectrum of time-scales, the matrix $\Omega$ and the sources $\vec S$, we now consider the dynamics in the opposite limit, namely $r\approx 1 \rightarrow \xi\ll 1$. In this case, we approximate the inverse $(\hat I+\xi \bar \Omega W)^{-1}$ with $\approx \hat I-\xi \Bar \Omega W$. We can thus write:
\begin{eqnarray}
  \frac{d  w_i}{d t} &\approx&\alpha w_i-\frac{1}{\beta} \left( (I-\xi\ \bar\Omega W)\bar \Omega \bar S\right)_i\nonumber \\
 &=&\alpha w_i-\frac{ \xi}{\beta} \sum_{jkt} \bar \Omega_{ij} w_j \delta_{jk} \bar \Omega_{kt} \bar S_t + s_i\nonumber \\
 &=&\sum_j \left(\alpha \delta_{ij}- \frac{\xi}{\beta} \sum_{t} \bar \Omega_{ij} \bar \Omega_{jt} \bar S_t \right) w_j +s_i\nonumber \\
 &\equiv& \sum_{j} O_{ij} w_j+s_i,
\label{eq:dynnndc}
\end{eqnarray}
where $s_i=\frac{\xi}{\beta} \sum_j \bar \Omega_{ij} \bar S_j$. The solution of this equation is given by:
\begin{equation}
\vec w(t)=e^{O t} \left({\vec w}^0+\int_{t_0}^t e^{-O \tilde t} \vec s(\tilde t) d\tilde t \right).
\end{equation}
In the case of a dc-controlled memristive network, e.g. $\vec s(t)=\vec s$, inevitably the memristors will reach their boundary values, $1$ or $0$. If we, however, focus on the short dynamics of memristors, it is interesting to study the distribution of eigenvalues of the matrix $O$. For the case of random matrices $O$ with passive components,  the distribution of eigenvalues is given in Fig. \ref{fig:eigs}. The first observation is that the distribution is symmetrical and that for $L/M\rightarrow 1$ the distribution flattens out. Given that we have randomly generated voltages on the memristors in $[V/2,V/2]$ with $V=100$, we observe that in this limit $\tilde \lambda$ of  \eqref{eq:approx} can be roughly assumed to be $V/2$. In all the random graph classes studied here, for high density of loops the matrix $\Omega$ becomes non-sparse, and thus a dynamical mixing behavior emerges. This is a again universal property which seems to be the underlying reason for observing the slow relaxation behavior for DC controlled circuits.

The emergence of a scale-free thermalization of the mean internal parameter shows that disordered memristive networks exhibit  an \textit{aging} phenomenon, 
typical of glassy materials \cite{sk,Vincent}. Although being dramatically different, an analogy to the Sherrington-Kirkpatrick model stems naturally in our analysis: the circuit graph generated at random induces a random projector $\Omega$, which in turn induces a random coupling matrix $(I+\xi\ \Omega)^{-1}$. Therefore, in this simple model there is no notion of distance, and thus there is strong non-locality. We cannot thus discuss of spatial  correlations which fall off as a power law. 

The slow relaxation behavior observed is compatible with the experimental results obtained in the case of \cite{Stieg12}, where it was observed that the power spectrum scales with a power law behavior with exponent $\approx1.34$; it was also noted that such scaling is an order of magnitude larger than the noise, and that the effect was due to a collective network effect. In the low-memory regime described above, such exponent can be easily calculated to be equal to $2$. Thus, although our analysis fails to explain the observed exponent, it must be noted that we are considering the linear regime only, and that the memristors we study are of the simplest linear type.

\subsection{AC approximate solution} For the case of AC forcing, using \eqref{eq:dynnn5} it is possible to go beyond the exponential approximation, and provide a solution in the approximation of diagonally dominant $\Omega$ matrices. For simplicity, here we derive the exact solution for the case of a single mesh with only one memristor with a voltage generator, but a full derivation is provided in the Supporting Information. We consider first the relation between the voltage, the current and a memristor:
\begin{eqnarray}
V(t)&=& R_{on}[1+\xi w(t)] I \nonumber \\
&=& \beta [1+\xi w(t)] \frac{d w}{dt} \nonumber \\
&=& \beta \frac{d}{dt}[w(t)+\frac{\xi}{2} w(t)^2].
\label{eq:start1}
\end{eqnarray}
If we define the flux $\Delta \Phi(t)=\int dt V(t)$, and integrate both sides of \eqref{eq:start1}, we obtain: 
\begin{equation}
\Delta \Phi(t)=\beta [w(t)+\frac{\xi}{2} w(t)^2]-K_0,
\end{equation}
where $K_0=\beta [w(0)+\frac{\xi}{2} w(0)^2]$. Inverting for $w(t)$, one obtains:
\begin{equation}
w(t)=\frac{\beta - \sqrt{ \beta^2+2 \xi \beta  (\Delta \Phi(t)+K_0)}}{\beta \xi},
\label{eq:solone}
\end{equation}
and where we chose the solution sign for which $w(t)$ is always positive. In general,  we observe that $0 \leq w(t) \leq 1$. If one introduces a sinusoidal potential $V(t)=v_0 \sin(\omega t)$, $\Delta \Phi(t)=- \frac{v_0}{\omega} \cos(\omega t)$. It is easy to see that in this case $\lim_{\omega\rightarrow \infty} w(t)=w_0$, implying that for large frequencies memristors loose their memory properties, as expected.\\
The case of a network is a generalization of the above procedure. The full derivation is provided in the Supporting Information, but it is important to stress that such approximate solution applies when the matrix $\xi \bar \Omega W$ is diagonally dominant and if the dynamics is continuous, i.e., if no memristor effectively reaches the boundary values. The former requirement depends on the network topology, while the latter is a condition on the applied voltages, i.e., if the forcing is AC and the voltages are small enough. 

\begin{figure}
\includegraphics[scale=0.21]{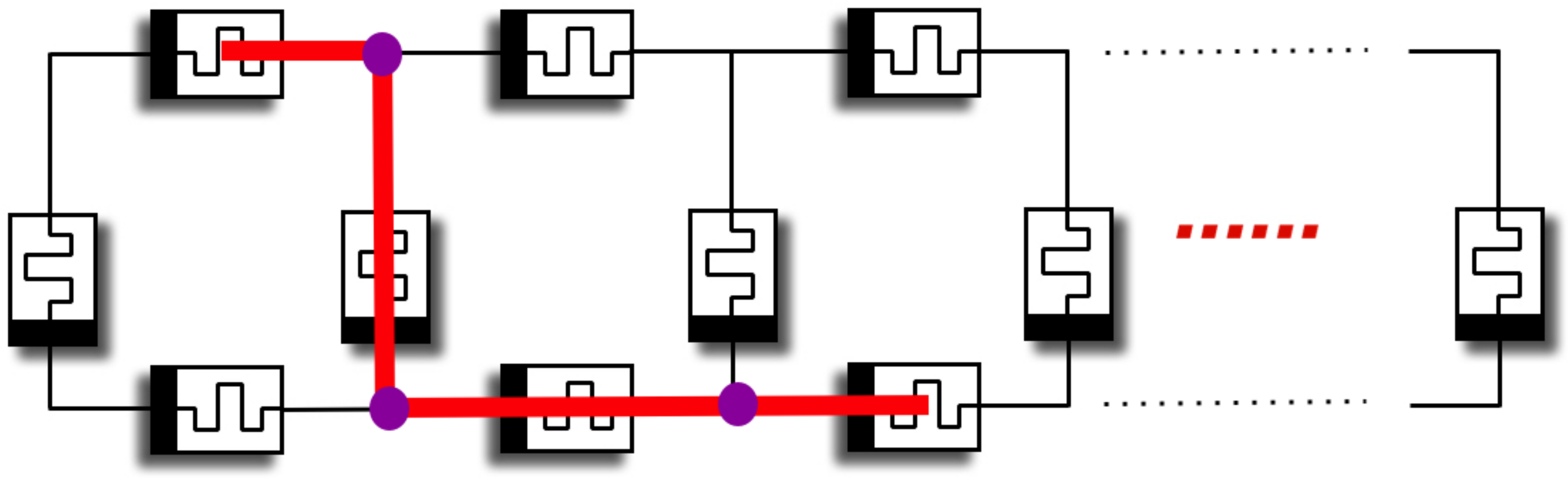}
\caption{Simple rectangular circuit and the definition of Hamming distance between the memristors, defined as the minimum number of nodes required to be traversed in order to reach the two memristors.}
 \label{fig:example}
 \end{figure}

Analogously to the case of a single memristor, for a network we define $\Delta \Phi_i\equiv \int_{t_0}^t \sum _j \bar\Omega _{\text{ij}} S_j(s)  ds$, and is given by:
\begin{eqnarray}
w_k(t)&=&\sum_{i} \bar \Omega _{ki}\frac{ \left(\sqrt{(1+\xi \sum_j \bar \Omega_{ij} w_j^0)^2 +\frac{2 \xi }{\beta}\Delta \Phi_i}-1 \right)}{  \xi }\nonumber \\
&+&\sum_i (I-\bar \Omega)_{k i} w^0_i,
\label{eq:sol0}
\end{eqnarray}
and can be obtained by means of quadrature.\\\ \\
We note that such solution exists as long as, given a set of sources identified as $\{ \omega_j \}$, the frequencies satisfy the condition:
\begin{equation}
(1+\xi \sum_j \bar \Omega_{ij}  w^0_j)^2 > \frac{2 \xi}{\beta} \sum_j \bar \Omega_{ij} \frac{1}{\omega_j}\ \ \ \forall i,
\end{equation}
which is the requirement for the quadrature method we employed to work.
In order to check the validity of our solution, we compare it to a numerically integrated solution of the differential equation for a simple network configurations. This solution is exact for the trivial case of parallel memristors, for which $\Omega$ is exactly diagonal. 
We test the validity of \eqref{eq:sol0} in the case of a simple network, which we show in Fig. \ref{fig:example}.
For such case, we first arrange the matrix $\Omega$ which characterizes the circuit in such a way that two memristors are further away according to the Hamming distance on the graph. 

\begin{figure}[ht!]
\centering
\includegraphics[scale=0.28]{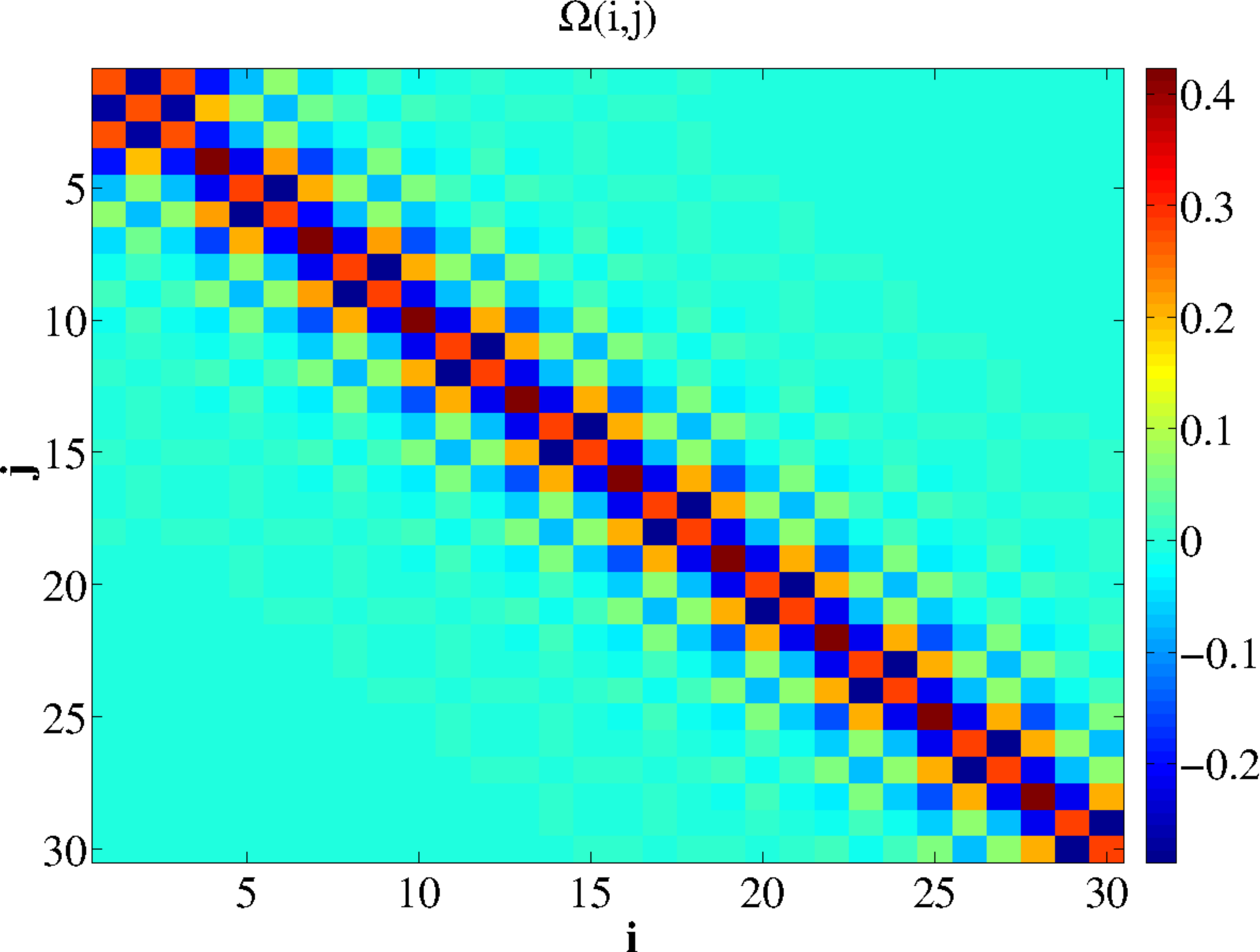}\\
\includegraphics[scale=0.26]{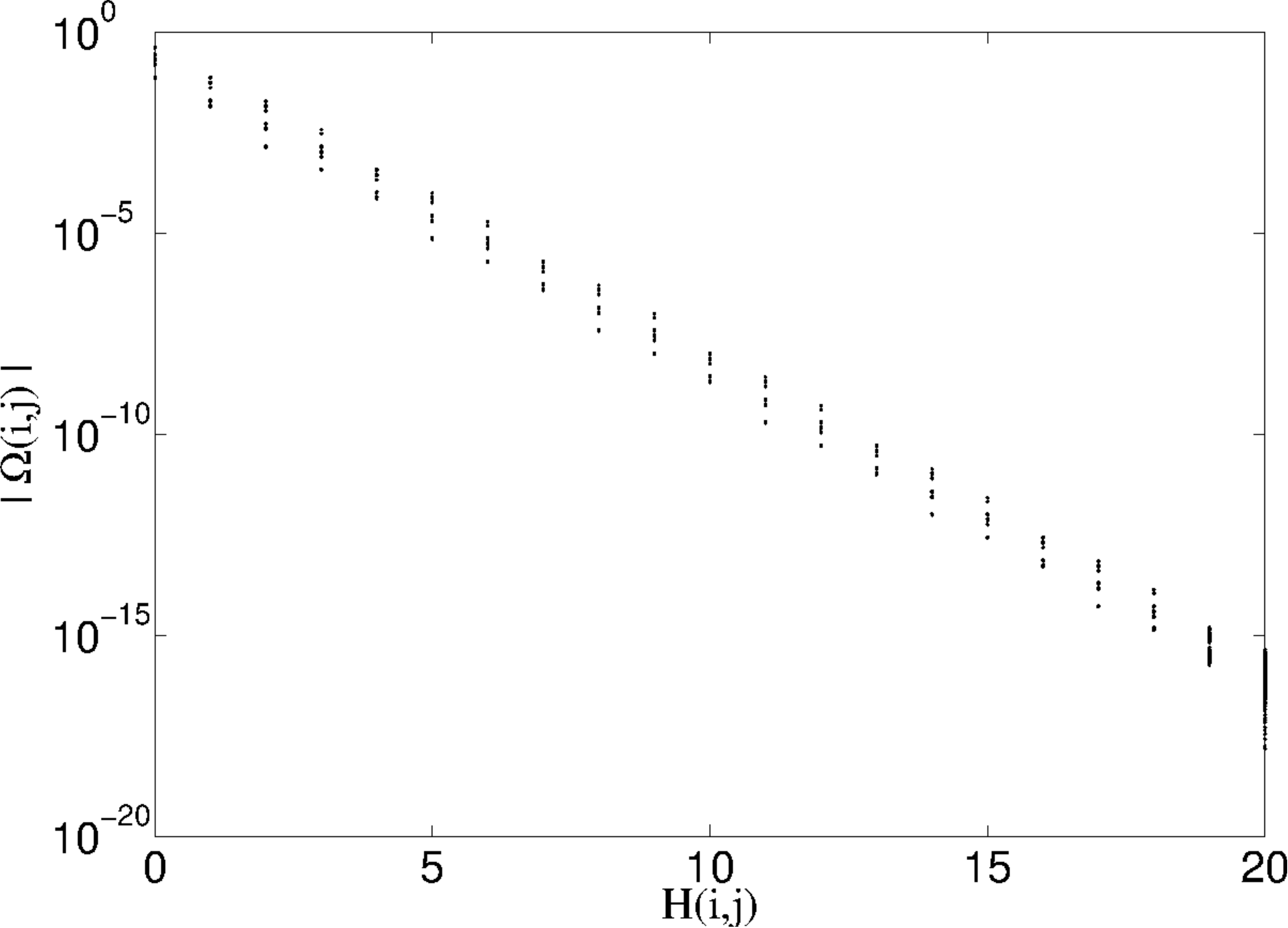}
\caption{Elements of the projector operator $\Omega$ for the circuit in Fig. \ref{fig:example}, sorted according to the Hamming distance $H(i,j)$ between the memristors $(i,j)$. We observe that $\Omega$ (top) is a diagonally dominant operator, with strength which decays exponentially in absolute value as a function of the Hamming distance (bottom). The observed degeneracy is due to the fact that there are many memristors with the same Hamming distance.}
\label{fig:locality}
\end{figure}

\begin{figure}[!ht]
 \centering
\includegraphics[scale=0.28]{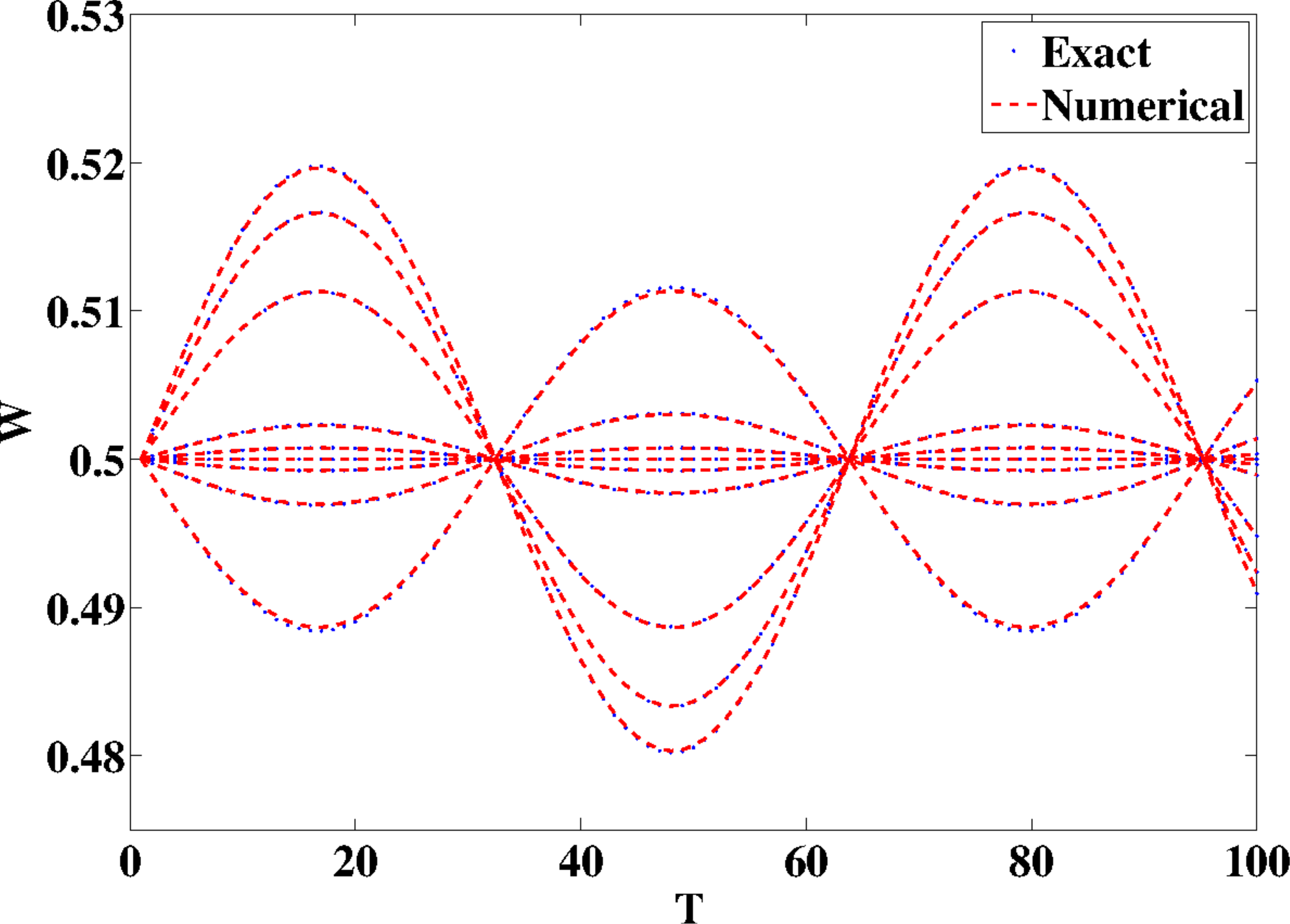}\\
\includegraphics[scale=0.28]{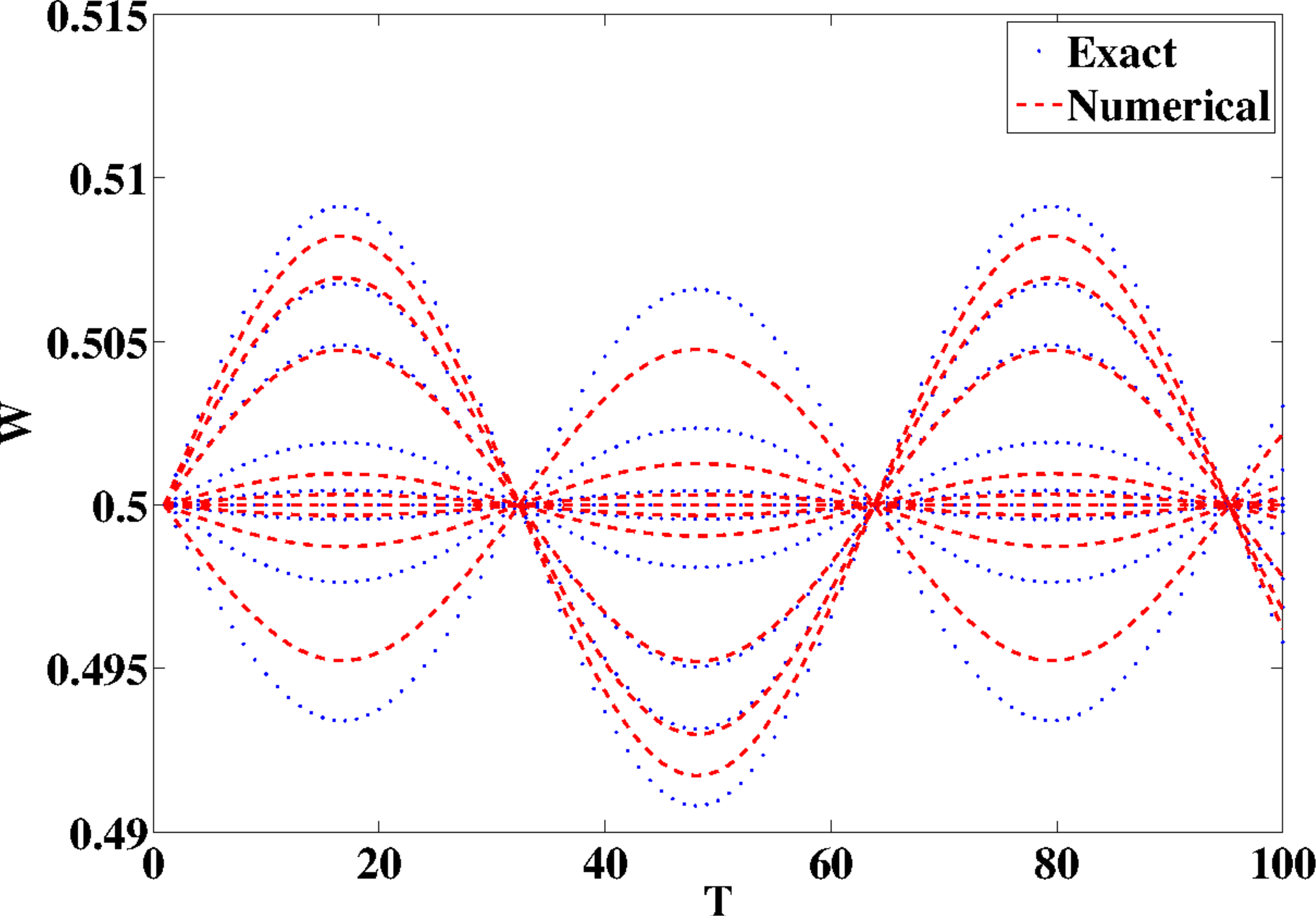}
\caption{Analytical approximation versus numerical integrations of the \eqref{eq:dynnn5} as compared to the approximate solution of \eqref{eq:sol0}. We choose $R_{on}=\beta=100$ and $r=1000$, with $\omega=30$. \textit{Top:} Numerical versus exact solution of the circuit of Fig. \ref{fig:example} for $\xi=0.1$. \textit{Bottom:} Numerical versus exact solution of the circuit of Fig. \ref{fig:example} for $\xi=3$. The deviation from the exact solution is due to the strength of the off-diagonal terms of the matrix $I+\xi \Omega W$ for large values of $\xi$. }
 \label{fig:solcomp}
 \end{figure}

The matrix $\Omega$ with this labeling and ordering is shown in Fig. \ref{fig:locality} (top). We note that graphs that are quasi-local imply diagonally-dominant matrices $\Omega$. Interactions strengths, in this case, decay exponentially as a function of the Hamming distance, as it is shown in Fig. \ref{fig:locality} bottom. This is a specific example in which it is shown that even for circuits which are regular, non-local spatial correlations emerge, although these being exponentially weighted.
Such circuit example serves the purpose of graphs for which the matrix $\Omega W$ is diagonally dominant, due to the fact that $W$ is diagonal. In Fig. \ref{fig:solcomp} we thus compare the approximate solution of \eqref{eq:sol0} to the one obtained numerically, both for the case of small values of $\xi$ and for the case of larger values. We observe an excellent agreement with the numerical curves for $\xi\ll 1$. The limit of such approximation is however clear for larger values of $\xi$, with the exact solution deviating for a few memristors from the numerical solution.

\section{Discussion}
In conclusion, we have derived a matrix differential equation for the evolution of the internal memory states of linear current-driven memristors in a circuit. The differential equation is general and solves the circuit constraints based upon the graph theoretical derivation of the current configuration, using the formalism of spanning trees and fundamental loops in networks. Such equation establishes a very clear connection between the theory of projector operators and the dynamics of linear memristive circuits. In particular, we have found that the network dynamics of the internal memory of the circuit can be constructed from only a single matrix, the projector on the space of fundamental loops of the circuit, which contains the information on the network topology and the conservation laws. This shows that the internal memory of the memristors is insensitive to certain forcing modes which fall in the kernel of the projection operator. We believe such equation to be an important tool for obtaining a deeper understanding of the dynamics of the internal memory of memristive circuits. 

By focusing on the case of dc-controlled circuits, we have provided sufficient evidence that the average relaxation of the internal memory to the boundary values is far from exponential and ``universal'', namely topology-independent. We have shown this by generating random projection operators, but such result is consistent with simulations performed using memristive random circuits.
We have also given arguments that the slow relaxation is due to the superposition of memristive dynamics decaying with a broad range of time scales at least in the regime of ``shallow'' memory, i.e., when non-linearities are negligible. 

For the case of AC controlled circuits, we have shown that an approximate solution can be obtained, and which agrees with the numerical solution in the limit of the approximation of low memory and for diagonally dominant projector operators. This also provides direct evidence of the usefulness of the derived equations, and shows in a specific sense how the graph topology and the constraints enter the dynamics of the internal memory. We have also studied the projector operator in the case of a simple graph with a local structure. We have shown that although the graph exhibits a local structure, a certain amount of non-local spatial correlations emerges due to the circuit constraints. This non-local correlation is found to be bounded by an exponential function in the Hamming distance on the graph.

These results reveal the rich dynamics of complex networks with memory and establishes a new research direction in memristive circuits. We believe in fact that a similar equation can be derived for other types of memristors, allowing a deeper understanding of the relation between memory (time non-locality) and topology of the graph.


\clearpage
\begin{centering}
\textbf{Acknowledgements}
\end{centering}
Two of us (FLT and MD) acknowledge partial support from the Center for Memory Recording Research at UCSD. FC thanks Invenia Labs for support and Thomas Fink for reading the manuscript and providing useful suggestions.

\clearpage
\appendix

\section{Supporting Information}
\subsection{Formal solution of linear circuits} \label{sec:derivationstart}
In this section we recall the basics of graph theory used to derive \eqref{eq:dynnn6} and provide the notation used throughout the main text \cite{nilsson,bollobas}. We start by considering a graph $G$ with $N$ nodes and $M$ edges which describes the topology of a resistive circuit. As it is standard practice, we then introduce an orientation $\mathcal O$ for the currents circulating on the graph, which has $2^m$ possible configurations, with $m$ being the number of edges (or arcs). From the point of view of graph theory, the graph representing the circuit must be connected, and the degree of each node $i$ satisfies $d_i>2$. For each node, we can introduce a potential vector $p_\alpha$, and for each edge a current $i_k$, where we use latin indices for the edges, and greek indices for the nodes; greek indices with tildes represent instead cycles on the graph. Given an orientation $\mathcal O$, we can introduce two matrices: $B^{\mathcal O}_{\alpha k}$, which is a matrix of size $N \times M$, and a cycle matrix $A^{\mathcal O}_{\tilde \xi m}$, which is of size $C \times M$, where $C$ is the number of cycles of the graph. From now on, we will suppress the orientation apex for simplicity. A valid current configuration is one in which $\sum_{j=1}^M B_{\alpha j} i_j=B \vec i=0$, which represents the Kirchhoff Current Law (KCL). In order for $B$ to have the linear independence of the rows, one row has to be removed, introducing thus the \textit{reduced incidence matrix}. In the following, we will thus consider only results derived with this matrix rather than the full one.

 Given a potential vector based on the nodes, the vector of voltages applied to each edge can be written as $\{\bar v\}_k=v_k= \sum_{\xi} B_{ \xi k}^t p_k$, where $t$ represents the matrix transpose. The Kirchhoff Voltage Law (KVL) can thus be written as $\sum_k A_{\tilde \xi k} v_k=0$, an equation which is simply saying in graph theory terms that the circuitation of the voltage on edges belonging to a cycle (or  \textit{mesh} in circuits) must be zero. This automatically implies that in general $B\cdot A^t=A\cdot B^t\equiv 0$. Analogously, this implies that $\vec i \cdot \vec v=0$, which represents the conservation of energy, or \textit{Tellegen's theorem} in circuits. 
 
 Let us now introduce a spanning tree $\mathcal T$ (\textit{co-chords}), and the set of edges of the graph not included in the tree as $\mathcal T$, \textit{ or chords}, are given by $\bar {\mathcal T}$. For each element of the chord $\bar {\mathcal T}$, we assign a cycle, called \textit{fundamental loop}. The number of fundamental loops is given by $L=M-N+1$. As a matter of fact, each current can be written as a superposition of the currents flowing in the fundamental loops, denoted with $j_{\tilde \xi}$, and one has that $\vec i= A^t \vec i_c$. In the basis in which we reorder the edges in the tree to come first, one can write $A=(A_{\mathcal T}, A_c)$, and since now $A_{c}$ corresponds to fundamental cycles, $A_{c}$ is the identity matrix, $A_{c}=I$. A similar rearrangement can be made also for the the incidence matrix $B$, and thus one has $(B_{\mathcal T}, B_{c})\cdot (A_{\mathcal T},\ I)^t=0$, which implies $A^t_{\mathcal T}=- B_{\mathcal T}^{-1} B_{c}$. We now note that since $B \vec i=0$, one has $B_{\mathcal T} \vec i_{\mathcal T}+B_{c} \vec i_{c}=0\ \rightarrow \vec i_{\mathcal T}=- B_{\mathcal T}^{-1} B_{c} \vec i_{c}=A_{\mathcal T}^t i_c$. Since $A_{c}=I$, this implies that $\vec i=(A_{\mathcal T}^t {\vec i}_c, {\vec i}_c)=A^t {\vec i}_c$. Since $A$ is derived from the \textit{reduced} incidence matrix, this is called \textit{reduced loop matrix}.

If we now write the equation for the circuit, i.e., $\vec v=R \vec i+ \vec S(t)$, we note that applying the operator $A$ on the left, we obtain the identity $A \vec v=0 =A R \vec i+A \vec S(t)$. We now use ${\vec i}=A^t {\vec i}_c$, and obtain $(A R A^t) {\vec i}_c=- A \vec S_{0}(t)$. If we now write the solution of the current, we obtain 
\begin{equation}
\vec i=A^t {\vec i}_c=- A^t(A R A^t)^{-1} A \vec S(t). 
\label{eq:init0}
\end{equation}
which is the starting point of this paper. We stress that since $A$ is derived from a reduced incidence matrix, then $AR A^t$ is always invertible for non-zero resistances.\ \\

\subsection{Derivation of the main dynamical equation}\label{sec:derivation}
The starting point of the derivation is  \eqref{eq:init0}. We consider the convention in which $w=0$ corresponds to $R_{on}$ and $w=1$ to $R_{off}$, which is a memristor with opposite polarity to the one introduced in  \cite{strukov08}.
First of all, let us first say that it is easy to parametrize the presence of active components. In this case, one can simply introduce negative resistances in  \eqref{eq:init0}, for instance introducing a matrix $P=\text{diag}(\pm 1,\cdots,\pm1)$, where $+1$ are assigned to passive components, while $-1$ to active components. It is easy to see that $P$ satisfies the property $P^2=I$. The resistances are thus encoded in the matrix $\tilde R=P R=R P$, and \eqref{eq:init0} simply becomes
\begin{eqnarray}
\vec i&=&- A^t \left( A \tilde R A^t\right)^{-1} A \vec S(t) \nonumber \\
&=&- A^t \left( A P R A^t\right)^{-1} A \vec S(t) \nonumber \\
&=&- A^t \left( \bar A  R A^t\right)^{-1} A \vec S(t). \nonumber
\label{eq:init2}
\end{eqnarray}

As in the main text, we define the matrix $\bar A=A P$ and  we also use the dynamical properties of the memristors, $\frac{d}{dt} w_j={\mathcal J_j} (R_{on}/\beta)\  i_j +\alpha w_j$, with ${\mathcal J}_j$ representing the polarity of the memristor. The goal of this section is to derive a dynamical equation which is in terms of projectors only. For this purpose, we  use the Woodbury identity to write the equation in terms of projector only,
\begin{equation}
(Q+ U C V)^{-1}=Q^{-1}\left(I-U (C^{-1}+V Q^{-1} U)^{-1} V Q^{-1} \right)
\label{eq:wood}
\end{equation}
where $Q$ and $C$ are square matrices of size $n$ and $k$ respectively, and $V$ and $U$ are rectangular matrices, and which is valid as long as $Q$ and $C$ have inverses.

We first introduce the parameter $\xi=r-1$ which, as it will become clear soon, can be thought of as the amount of nonlinearity present in the system. 
Using  \eqref{eq:wood}, we are thus able to rewrite the inverse $(\bar A R A^t)^{-1}$, obtaining:
\begin{eqnarray}
(\bar A A^t + \xi\ \bar A W A^t)^{-1}= \nonumber \\
(\bar A A^t)^{-1}\left(I-\bar A \left(\frac{W^{-1}}{\xi}+\bar \Omega\right)^{-1} A^t(\bar A A^t)^{-1} \right)
\end{eqnarray}

where we introduced the operator $\bar \Omega\equiv A^t (\bar A A^t)^{-1} \bar A$. It is important to say that the operator $\bar A A$ is invertible, as we are considering the reduced loop matrix, and $\mathcal J$ is full rank by construction. Using  the result of  \eqref{eq:finalinverse} for the inverse $(\bar \Omega+B)^{-1}$ and derived in Sec.~\ref{sec:inver},
we can now obtain the final equation:
\begin{eqnarray}
\frac{\beta}{R_{on}} \frac{d \vec W}{d t}& =&-\frac{\mathcal J}{R_{on}}  A^t (\bar A A^t)^{-1}  \bar A  \left(I-\xi\  W \right)A^t (\bar A A^t)^{-1}  AS(\tilde t)  \nonumber \\
&+&\frac{\beta}{R_{on}} \alpha   \vec W  \nonumber \\
&+&\frac{\xi ^2}{R_{on}}\ \mathcal J  \bar \Omega (I+\xi\ W \bar \Omega)^{-1} W \bar \Omega W A^t (\bar A A^t)^{-1}  A  \vec S,\nonumber 
\label{eq:dynnnact}
\end{eqnarray}
where we introduced $\mathcal J=\text{diag}({\mathcal J}_i)$. If we now use the identity $P^2=I$, we can write $\vec S=P^2 \vec S=P \bar S$, and  obtain the equation
\begin{eqnarray}
\frac{\beta}{R_{on}}   \frac{d \vec W}{d t} &=&-\frac{1}{R_{on}} \mathcal J   \bar \Omega (I -\xi\ W )\bar \Omega \bar S( t) +\frac{\beta \alpha}{R_{on}}   \vec W\nonumber \\
 &+&\frac{1}{R_{on}} \xi^2 \mathcal J  \bar \Omega (I+\xi\ W \bar \Omega)^{-1} W\bar \Omega W \bar \Omega \bar S( t),\nonumber \\
\label{eq:dynnnactfin}
\end{eqnarray}
in which we used the fact that $[W, \bar S]=0$ since the two matrices are diagonal and $\mathcal J=I$. Finally, we use the fact that $\bar \Omega (I+\xi W \bar \Omega)^{-1}=\bar \Omega\sum_{k=0}^\infty (-\xi W \bar \Omega)^k=\sum_{k=0}^\infty (-\xi\ \bar \Omega W )^k\bar \Omega= (I+\xi  \bar \Omega W)^{-1} \bar \Omega$, and obtain the final result shown in  \eqref{eq:dynnn6}.
Using the matrix Taylor expansion in $\xi=r-1$, we can finally write the equation:
\begin{eqnarray}
 \frac{d \vec W}{d t}&=&\alpha \vec W -\frac{1}{\beta}  \mathcal J  (I+\xi\ \bar\Omega W)^{-1} \bar \Omega \bar S( t), 
 \label{eq:dynnn6}
\end{eqnarray}
which is our final result. 
We note that $\bar \Omega=A^t (\bar A A^t)^{-1}  \bar A$ is the most general description of a non-orthogonal projection operator. In the matricial limit $P\rightarrow I$, then $\bar A\rightarrow A$ and thus the projector becomes orthogonal again.

It is also easy to see that for $P=\pm I$, then $\bar A=\pm A$. As such, $\bar \Omega$ is indeed an invariant under this symmetry. In fact, $(\bar A A^t)^{-1}=\pm (A A^t)^{-1}$, and thus $\Omega$ is invariant. However, $\bar S\rightarrow \pm S$, which means that this is simply a change of current flow. Although currents in a circuit are defined up to a change of direction, this implies that the circuit is not invariant under this symmetry. 

The above result can, in principle, be extended with little effort to the case of voltage-controlled memristors. Let us assume that the equation for the evolution of internal memory is of the type:
\begin{equation}
\frac{d \vec W}{dt}=\rho \vec V.
\end{equation}
In this case, we can write:
\begin{eqnarray}
\frac{1}{\rho}\frac{d \vec W}{dt}&=&\vec V = R \vec i,
\end{eqnarray}
and using the equations we derived, we obtain:
\begin{eqnarray}
\frac{d \vec W}{dt} &=& \rho R_{on} (I+\xi W) \vec i \nonumber \\
&=& \rho (I+\xi W) \mathcal J  (I+\xi\ \bar\Omega W)^{-1} \bar \Omega \bar S( t),
\end{eqnarray}
which is the extension of the equation above to the case of voltage-controlled memristors.

\subsection{Matrix Inverse} \label{sec:inver}
Let us now prove an inversion equation, where we assume that $B$ and $\Omega+B$ are invertible:
\begin{equation}
(\Omega+B)^{-1}=B^{-1}+X
\end{equation}
and aim to find the matrix $X$. By definition, we have:
\begin{eqnarray}
I&=&(B^{-1}+X)(\Omega+B)\nonumber \\
&=&\left(B^{-1} \Omega+X( \Omega+B)+B B^{-1}\right)\nonumber \\
&=&\left(B^{-1} \Omega+X( \Omega+B)+I\right).
\end{eqnarray}
Using this identity, we can find the matrix $X$ by inversion:
\begin{eqnarray}
X&=&- B^{-1}\Omega(\Omega+B)^{-1} \nonumber \\
&=&- B^{-1}\Omega(B^{-1}+X) \nonumber \\
&=&- B^{-1}\Omega B^{-1}-B^{-1}\Omega X,
\end{eqnarray}
which can be rewritten as:
\begin{equation}
X=-(I+B^{-1} \Omega)^{-1} B^{-1} \Omega B^{-1}.
\end{equation}
This implies the following inversion formula:
\begin{equation}
(\Omega+B)^{-1}=B^{-1}-(I+B^{-1} \Omega)^{-1} B^{-1} \Omega B^{-1},
\label{eq:finalinverse}
\end{equation}
which is the identity we use to reach the final equation, and requires only the invertibility of the matrices $B$ and $\Omega+B$, but not the invertibility of $\Omega$. Such identity is important since in our case $\Omega$ represents a projector operator, meanwhile $B$ represents $W^{-1}/(r-1)$. Starting from the identity $I=(\Omega+B)(B^{-1}+X)$, we obtain the identity:
\begin{equation}
(\Omega+B)^{-1}=B^{-1}-B^{-1} \Omega B^{-1}(I+\Omega B^{-1} )^{-1} ,
\label{eq:finalinverse}
\end{equation}
which implies that: 
\begin{equation}
(I+B^{-1} \Omega)^{-1} B^{-1} \Omega B^{-1}=B^{-1} \Omega B^{-1}(I+\Omega B^{-1} )^{-1}.
\label{eq:finalinverse}
\end{equation}
This is the result we used in order to derive the differential equation for the internal memory.

\subsection{Finite size effects}\label{sec:feat}
\begin{figure}
\includegraphics[scale=0.33]{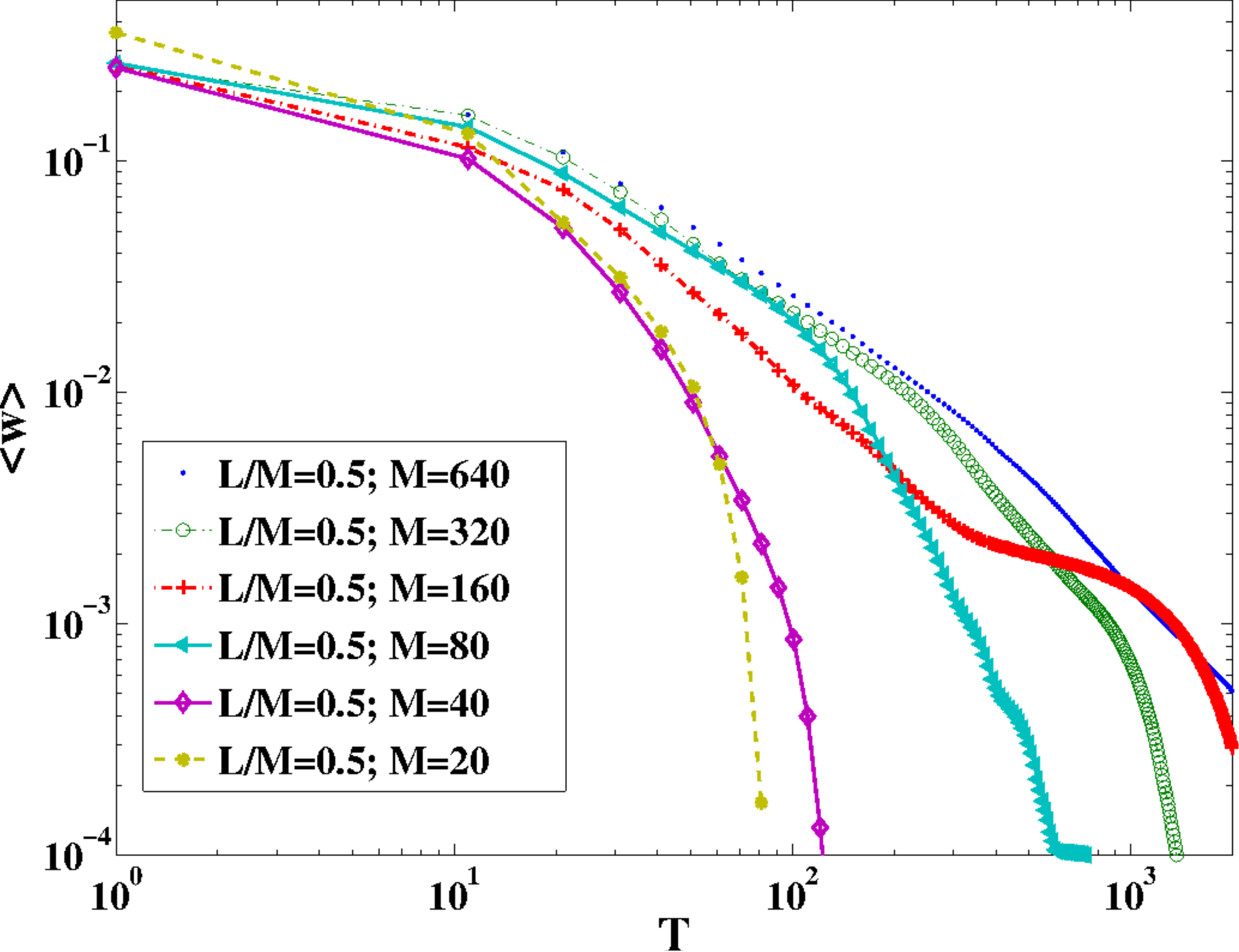}
\caption{ Average relaxations for fixed ratios $L/M=0.5$, with $M=20,40,80,160,320,640$ and initial memristors set at $w_i=1$. These are not averaged over many quenched dynamics, but single simulations. We observe that for increasing values of $M$, the curves suppress the fluctuations and converge to the slow dynamics phenomenon for the average memory.
}
\label{fig:nmfixed}
\end{figure}

We now mention some observed  factors influencing the quality of the power law.
The number of memristors controls the emergence of the power law decay. Numerically, we observe that $M\approx 100$ is enough to have a faithful power law. 
In Fig. \ref{fig:nmfixed} we can see that when $L/M$ is fixed and $M\rightarrow \infty$, the limit $M$ can be considered as a thermodynamic limit, i.e., the power law emerges for large $M$.

\subsection{AC approximate solution}
In this section we derive the approximate solution for the case of AC forcing of the main equation derived in the text. For a single memristor (or single mesh) it is in fact possible to derive an exact equation for the memory, and this applies also to the simpler case of many memristors in parallel, for which $\Omega$ is a diagonal matrix.
It is thus clear that for a class of graphs for which $\Omega$ is diagonal such quadrature can be obtained in the case in which the memristors will not reach the boundary values, thus in the approximation of differentiable dynamics.

It is thus important that we understand first in which sense it is possible to approximate a matrix by a diagonal one.
As it turns out, this statement depends on the type of functional one aims to minimize. Specifically,  in which sense is it possible to say that
\begin{equation}
Z=\bar \Omega W\approx Y,
\end{equation}
where $Y$ is a diagonal matrix? 
Such approximation is valid for the case in which $Z$ and $Y$ are acting on a vector, as we will show shortly.  We demand that, given any vector $\vec v$, 
$$Z \vec v-Y \vec v \approx 0$$ 
is true point-wise for each element of the resulting vector, and thus obtain
\begin{eqnarray}
\sum_{j,k} \Omega_{ik} \delta_{kj} w_j v_j=\sum_{j} \Omega_{ij} w_j v_j \approx \sum_j y_i \delta_{ij} v_j,
\end{eqnarray}
thus, for any element $v_j$, one has:
\begin{equation}
\sum_{j}(\Omega_{ij} w_j-y_i \delta_{ij})v_j=0\ \ \ \forall \vec v,
\end{equation}
and thus we require that
$\Omega_{ij} w_j-y_i \delta_{ij}\approx0$.
Summing over $j$ we obtain:
\begin{equation}
y_i=\sum_j \Omega_{ij}w_j.
\end{equation}
Once we have derived the diagonal matrix which approximate $\bar \Omega W$,  the derivation follows similar steps to the case of a single memristor. We first multiply \eqref{eq:dynnn6} by $\bar \Omega$, noticing that the rhs of this equation is invariant under this transformation, and subsequently by $(I+\xi \bar \Omega W)$, obtaining
\begin{eqnarray}
\beta (I+\xi\ \bar \Omega W) \bar \Omega \frac{d \vec W}{d t} &=&- \bar \Omega \bar S(t).
\label{eq:dynnn2}
\end{eqnarray}
Since $\bar \Omega$ is a projector, it is necessary to account at the end for this step. We now use the diagonal approximation of the matrix $\bar \Omega W$ applied on the vector $\bar \Omega \frac{d \vec W}{dt}$,
we can now repeat the steps of the single memristor case for each single element, with the key difference that now the memory has to be projected on the loop space.
We introduce the variable $\vec y=\bar \Omega \vec w$, and after having introduced $Y=\text{diag}(\vec y)$, obtain:
\begin{eqnarray}
\beta (I+\xi Y)  \frac{d \vec Y}{d t} &=&- \bar \Omega \bar S( t).
\label{eq:dynnn3}
\end{eqnarray}
Since $Y$ is a diagonal matrix, $Y \frac{d \vec Y}{dt}$ acts on each element of the vector as $\frac{1}{2}\frac{d}{dt} y_i^2$.
We thus note that we can write the matrix equation again as vectorial equation, and for each component write:
\begin{eqnarray}
 \beta \frac{d}{dt} (y_i+\frac{\xi}{2} y_i^2)&=&-   (\bar \Omega \bar S(t))_i\ \text{and thus} \nonumber \\
 \beta (y_i+\frac{\xi}{2} y_i^2)&=&-\int_{t_0}^t (\bar \Omega \bar S(s))_i ds\ +\beta (y^0_i+\frac{\xi}{2} (y^0_i)^2) \nonumber \\
&=&-\int_{t_0}^t (\bar \Omega \bar S(s))_i ds\ \nonumber \\
&+&\frac{\beta}{2 \xi } \left((1+\xi \sum_j \bar \Omega_{ij} w^0_j)^2-1\right) \nonumber \\
\label{eq:dynnn4}
\end{eqnarray}

In the case in which the sources are AC controlled with $\omega\gg1$, the memory elements will oscillate without ever reaching the boundaries. In this case, the solution we provided fully describes the dynamics. In order to see this, we write explicitly the solution as a function of $\Omega$. We parametrize $\vec S$ as $ S_i(t)=v_i \cos(\omega_i t+\phi_i)$. Thus, we have
\begin{equation}
\int_{t_0}^t S_i(s) ds= \frac{v_i}{\omega_i} \left(\sin(\omega_i t+\phi_i)-\sin(\omega_i t_0+\phi_i)\right). 
\end{equation}
We now impose that the pseudo-inverse of the matrix $\Omega$ satisfies the initial condition of the differential equation. This implies the addition of a term $(I-\Omega)k(t)$ for an arbitrary vector $k(t)$. However, the vector $k$ can be fixed at time $t=0$ by imposing the initial condition. In this case, $w^0=w(t=0)=\Omega \Omega w^0+(I-\Omega) w^0=w^0$.
If we define $z_i^0=(1+\xi \sum_j \bar \Omega_{ij} w_j^0)^2$, 
and invert the quadratic equation, we obtain the final  \eqref{eq:sol} below:

\begin{eqnarray}
w_k(t)&=&\sum_{i} \bar \Omega _{ki}\frac{ \left(\sqrt{z_i^0 +\frac{2 \xi }{\beta }\sum_j \bar \Omega_{ij} \int_{t_0}^t S_j(s) ds}-1 \right)}{  \xi }\nonumber \\
&+&\sum_{i}(I-\bar \Omega)_{ki}w_j^0 ,
\label{eq:sol}
\end{eqnarray}
which is the equation found in the main body of the paper.

\end{document}